\documentclass[twocolumn,aps,prb,10pt,superscriptaddress,longbibliography]{revtex4-2}
\usepackage{amsmath}
\usepackage{bm}
\usepackage{glossaries}
\usepackage{graphicx}
\usepackage{multirow}
\usepackage{textcomp}
\usepackage{threeparttable}
\usepackage[dvipsnames,svgnames]{xcolor}
\usepackage{hyperref}
\hypersetup{
pdfnewwindow=true,      
colorlinks=true,        
linkcolor=Blue,         
citecolor=Blue,         
filecolor=Blue,         
urlcolor=Blue           
}
\usepackage{siunitx}

\newacronym{nep}{NEP}{neuroevolution potential}
\newacronym{mlp}{MLP}{machine-learned potential}
\newacronym{nn}{NN}{neural network}
\newacronym{snes}{SNES}{separable natural evolution strategy}
\newacronym{rmse}{RMSE}{root mean square error}
\newacronym{mae}{MAE}{mean absolute error}
\newacronym{pes}{PES}{potential energy surface}
\newacronym{stm}{STM}{scanning tunneling microscope}
\newacronym{afm}{AFM}{atomic force microscopy}
\newacronym{dw}{DW}{domain wall}

\newacronym{dft}{DFT}{density functional theory}
\newacronym{md}{MD}{molecular dynamics}
\newacronym{nemd}{NEMD}{nonequilibrium MD}
\newacronym{aemd}{AEMD}{approach-to-equilibrium MD}
\newacronym{sed}{SED}{spectral energy density}
\newacronym{pca}{PCA}{principal component analysis}
\newacronym{mbd}{MBD}{many-body dispersion}
\newacronym{pbe}{PBE}{Perdew--Burke--Ernzerhof}
\newacronym{1d}{1D}{one-dimensional}
\newacronym{2d}{2D}{two-dimensional}
\newacronym{3d}{3D}{three-dimensional}
\newacronym{airebo}{AIREBO}{adaptive intermolecular reactive empirical bond order}
\newacronym{pdos}{PDOS}{phonon density of states}

\newacronym{itc}{ITC}{interfacial thermal conductance}
\newacronym{tc}{TC}{thermal conductivity}
\newacronym{lj}{LJ}{Lennard-Jones}
\newacronym{tdtr}{TDTR}{time-domain thermoreflectance}
\newacronym{ilp}{ILP}{interlayer potential}
\newacronym{shc}{SHC}{spectral heat current}
\newacronym{ta}{TA}{transverse acoustic}
\newacronym{la}{LA}{longitudinal acoustic}

\newacronym{gaa}{GAA}{gate-all-around}
\newacronym{fet}{FET}{field-effect transistor}
\newacronym{negf}{NEGF}{non-equilibrium Green’s function}
\newacronym{bte}{BTE}{Boltzmann transport equation}
\newacronym{bti}{BTI}{bias temperature instability}
\newacronym{hci}{HCI}{hot carrier injection}
\newacronym{hde}{HDE}{heat diffusion equations}
\newacronym{omat24}{OMat24}{Open Materials 2024}
\newacronym{al}{AL}{active learning}
\newacronym{mcmd}{MCMD}{Monte Carlo molecular dynamics}
\newacronym{ai}{AI}{artificial intelligence}
\newacronym{aimd}{AIMD}{ab-initio molecular dynamics}
\newacronym{gpu}{GPU}{graphics processing unit}
\newacronym{gpumd}{GPUMD}{Graphics Processing Units Molecular Dynamics}
\newacronym{sti}{STI}{shallow-trench isolation}
\newacronym{fps}{FPS}{farthest-point sampling}

\DeclareSIUnit\angstrom{\text{Å}}
\DeclareSIUnit{\atom}{atom}
\DeclareSIUnit{\step}{step}
\DeclareSIUnit{\atomstepsecond}{\atom\step\per\second}
\begin{document}

\title{Device-Scale Atomistic Simulations of Heat Transport in Advanced Field-Effect Transistors}


\author{Ke Xu}
\email{kickhsu@gmail.com}
\affiliation{College of Physical Science and Technology, Bohai University, Jinzhou, P. R. China}
\affiliation{Department of Electronic Engineering and Materials Science and Technology Research Center, The Chinese University of Hong Kong, Shatin, N.T., Hong Kong SAR, 999077, P. R. China}

\author{Gang Wang}
\affiliation{College of Physical Science and Technology, Bohai University, Jinzhou, P. R. China}

\author{Ting Liang}
\affiliation{Department of Electronic Engineering and Materials Science and Technology Research Center, The Chinese University of Hong Kong, Shatin, N.T., Hong Kong SAR, 999077, P. R. China}

\author{Yang Xiao}
\affiliation{College of Physical Science and Technology, Bohai University, Jinzhou, P. R. China}

\author{Dongliang Ding}
\affiliation{Department of Electronic Engineering and Materials Science and Technology Research Center, The Chinese University of Hong Kong, Shatin, N.T., Hong Kong SAR, 999077, P. R. China}

\author{Haichang Guo}
\affiliation{Department of Electronic Engineering and Materials Science and Technology Research Center, The Chinese University of Hong Kong, Shatin, N.T., Hong Kong SAR, 999077, P. R. China}

\author{Xiang Gao}
\affiliation{CAS Key Laboratory of Mechanical Behavior and Design of Materials, Department of Modern Mechanics, University of Science and Technology of China, Hefei, Anhui 230027, China}

\author{Lei Tong}
\affiliation{School of Integrated Circuits, Huazhong University of Science and Technology, Wuhan, Hubei 430074, China}

\author{Xi Wan}
\affiliation{Engineering Research Center of IoT Technology Applications (Ministry of Education), School of Integrated Circuits, Jiangnan University, Wuxi, Jiangsu 214122, P.R. China}

\author{Gang Zhang}
\email{gangzhang2006@gmail.com}
\affiliation{Yangtze Delta Region Academy in Jiaxing, Beijing Institute of Technology, JiaXing, 314019 China}

\author{Jianbin Xu}
\email{jbxu@ee.cuhk.edu.hk}
\affiliation{Department of Electronic Engineering and Materials Science and Technology Research Center, The Chinese University of Hong Kong, Shatin, N.T., Hong Kong SAR, 999077, P. R. China}

\date{\today}

\begin{abstract}

Self-heating in next-generation, high-power-density field-effect transistor limits performance and complicates fabrication, presenting significant challenges for accurate heat transport simulations. To address this, we introduce NEP-FET, a machine-learned framework for device-scale heat transport simulations of field-effect transistors. Built upon the neuroevolution potential, the model extends a subset of the OMat24 dataset through an active-learning workflow to generate a chemically diverse, interface-rich reference set. Coupled with the FETMOD structure generator module, NEP-FET can simulate realistic field-effect transistor geometries at sub-micrometer scales containing millions of atoms, and delivers atomistic predictions of temperature fields, per-atom heat flux, and thermal stress in device structures with high fidelity. This framework enables rapid estimation of device-level metrics, including heat-flux density and effective thermal conductivity. Our results reveal pronounced differences in temperature distribution between fin-type and gate-all-around transistor architectures. The framework closes a key gap in multiscale device modeling by combining near-quantum-mechanical accuracy with device-scale throughput, providing a systematic route to explore heat transport and thermo-mechanical coupling in advanced transistors.

\end{abstract}
\maketitle

\section{Introduction \label{section:Introduction}}

As device dimensions have been aggressively scaled in accordance with Moore’s law, transistor architectures have evolved from bipolar junction transistors to planar MOSFETs, followed by Fin\gls{fet} and \gls{gaa} transistors \cite{Bhol2022RPN,Kumar2025JWSL}. At nanometer scales, heat dissipation is governed almost entirely by conduction, rendering heat transport a primary bottleneck for further performance gains and, in advanced \gls{fet}, a key driver of self-heating and reliability degradation \cite{Chen2021nrp,WANG2023MT}.

In this regime, Fourier’s law fails to capture the underlying physics of heat conduction \cite{Chen2021nrp,Zheng2024PNAS,Beardo2025NPJCM}, motivating microscopic approaches to heat transport. {Ab-initio} calculations can provide interatomic force constants, Hamiltonians, and vibrational eigenmodes, which, when combined with the \gls{bte}, \gls{negf} or \gls{md} methods, yield detailed insight into phonon transport in low-dimensional and nano-structured materials \cite{Gabourie2024IEEEIEDM, Sheng2024IEEETED, Yue2024FR, Sheng2025IEEETED}. However, \gls{negf} methods scale poorly for large or complex systems and are less effective under strong anharmonicity or high-temperature conditions \cite{Schlunzen2020PRL}. In comparison, \gls{bte} solvers provide higher computational efficiency and bridge macroscopic Fourier models with atomistic treatments \cite{Guo2016JCP,Zhang2021ijhmt}, though their computational cost remains substantial for realistic \gls{3d} device geometries \cite{AlI2014IJTS,Shomali2017IJTS,Hossein2021PANS}. \gls{md} simulations, on the other hand, explicitly resolve atomic motion and naturally capture strong anharmonicity, enabling studies of diverse heat transport properties, ranging from size-dependent thermal conductivity \cite{Li2025TX,liang2023Mechanisms} to transport in low-dimensional materials \cite{Xu2019PRB,Li2019JCP}, and \gls{itc} \cite{Shao2015IJHMT,Bao2018esee,Gu2021JAP,Chen2022RMP,liang2025probing}. However, the accuracy of \gls{md} depends critically on the interatomic potential, and \gls{md} simulations remain computationally demanding for large, heterogeneous, or multi-component systems. It is therefore difficult to identify a single method that can simultaneously capture quantum effects, anharmonic scattering, and realistic device geometries, motivating the development of multiscale frameworks that combine atomistic accuracy with continuum-scale efficiency \cite{Prasnikar2024AIR}.

To integrate these complementary strengths, multiscale thermal frameworks have been advanced. Hu {et al.} introduced GiftBTE \cite{Hu2023JPCM}, an open-source nongray phonon \gls{bte} solver that achieves high computational efficiency and supports fully \gls{3d} simulations of real materials and devices. Building on this foundation, they proposed a parameter-free \gls{bte} formulation based on first-principles phonon properties \cite{Hu2024FR}, eliminating empirical parameters while maintaining quantitative agreement with experiments across diverse materials. By coupling first-principles nongray \gls{bte} with hydrodynamic heat equations (\gls{hde}), they realized multiscale thermal simulations for \gls{fet} \cite{Sheng2024IEEETED}. Gabourie {et al.} \cite{Gabourie2024IEEEIEDM} proposed a physics-grounded, \gls{ai}-accelerated multiscale pipeline that links atomic-level material realism to chip-scale temperature prediction. Their workflow employs \gls{dft}-level \gls{mlp} combined with \gls{md} to derive thermal conductivity, then uses Monte Carlo methods to obtain device-scale temperature maps, and finally applies \gls{ai} models to centimeter-scale and device-level temperature measurements. This approach reduces ad hoc calibrations and corrects drift-diffusion biases in hotspot locations and peak temperatures. Together, these multiscale approaches highlight the central role of \gls{md} in quantitative \gls{fet}-level thermal simulations.

\begin{figure*}[ht]
\centering
\includegraphics[width=1.4\columnwidth]{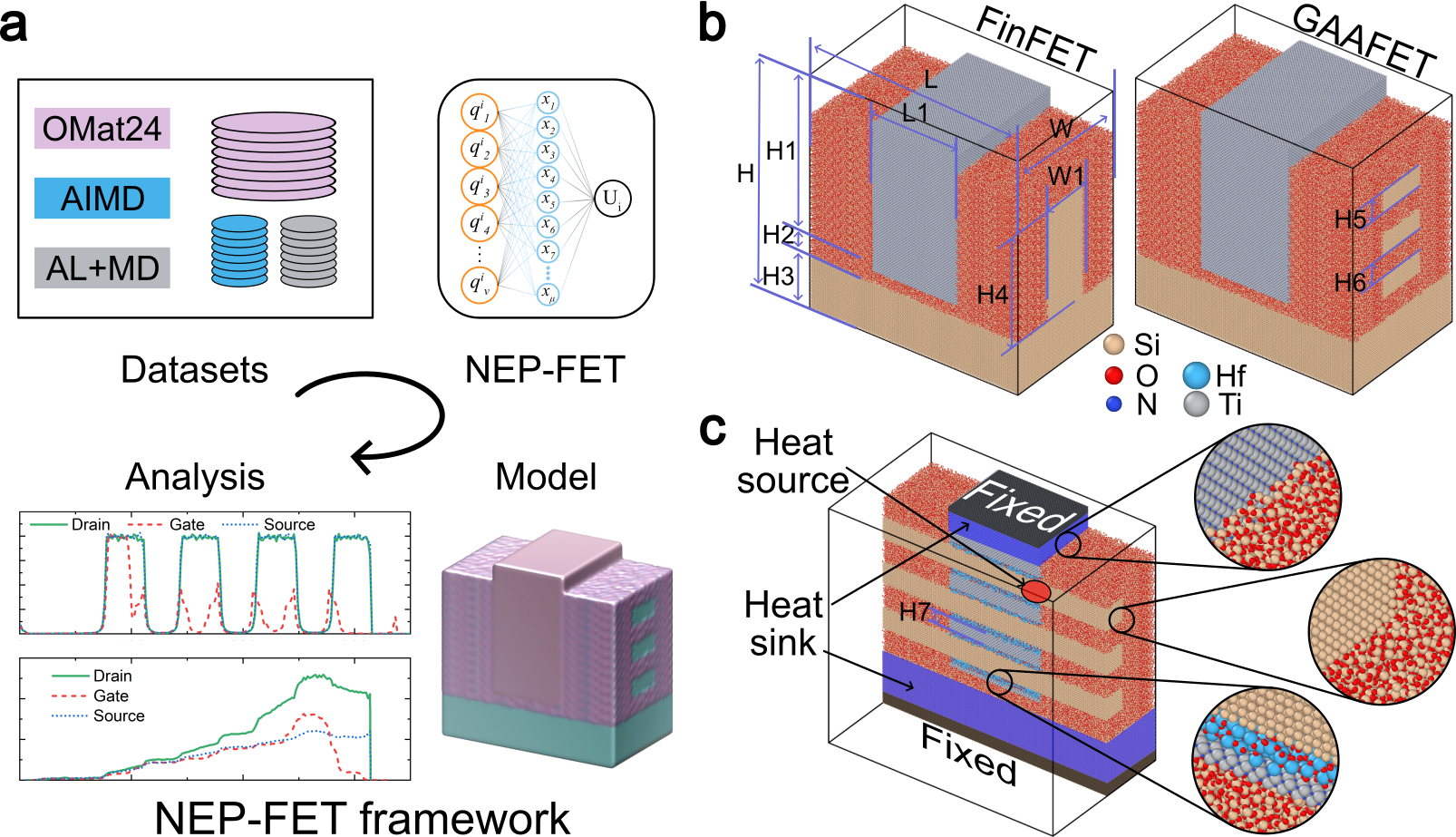}
\caption{
\textbf{Overview of the NEP-FET framework and representative \gls{fet} geometries.} 
(\textbf{a}) The NEP-FET framework is used for the simulation of device-level systems. The training set consists mainly of three components: (i) structures from OMat24 \cite{omat24}, (ii) configurations generated by \gls{aimd}, and (iii) configurations obtained via \gls{al}. These training sets are then processed and analyzed before being integrated into \gls{nep} for training. A realistic all-atom \gls{fet} model is then generated using the FETMOD module. Finally, device-level property calculations are performed using \gls{gpu}-accelerated \gls{md} software. (\textbf{b}) shows two typical \gls{fet} structures established using FETMOD module: Fin\gls{fet} and \gls{gaa}\gls{fet}, with typical device dimensions marked. Different colored regions are used to group the atomic models to accommodate subsequent \gls{md}-related calculations. (\textbf{c}) A magnified view shows the detailed features of the device structure.}
\label{fig:framework}
\end{figure*}

Recently, \gls{mlp} has emerged as a powerful solution to the accuracy-scale trade-off in \gls{md}. Foundation-style models such as MACE-MP-0 \cite{batatia2023foundation}, CHGNet \cite{deng2023chgnet}, M3GNet \cite{chen2022nuniversal}, DPA3 \cite{Zhang2025DPA3}, and NEP89 \cite{Ting2025NEP89} demonstrate broad chemical coverage, spanning multiple elements, phases, mixtures, and high-entropy alloys, while retaining near-quantum-mechanical accuracy. Combined with \gls{gpu}-accelerated \gls{md}, these models now enable simulations with thousands to millions of atoms on commodity hardware. For instance, \gls{nep} can deliver high fidelity for systems with tens of millions of atoms on a single \gls{gpu}, unlocking fully atomistic device-scale modeling \cite{Ting2025NEP89}.

Building on these advances, we first develop a specialized \gls{mlp} tailored for transistor devices, achieving near-quantum-mechanical accuracy in multi-million-atom simulations on a single GeForce RTX 4090 card. Our workflow reaches sub-micrometer spatial scales, enabling fully atomistic modeling of individual transistors. By considering the full-frequency phonon excitations intrinsic to \gls{md}, we compare heat-flux driving schemes and temperature-response characteristics in realistic transistor geometries, revealing the coupling between device architecture, heat generation, and nanoscale transport. This capability enables direct, quantum-mechanical-level investigation of self-heating phenomena in nanoscale transistors, bridging atomistic physics with device-level thermal behavior.


\section{Results \label{section:Results}}

\noindent\textbf{NEP-FET framework:} NEP-FET is a \gls{gpu}-accelerated simulation workflow for \gls{fet} built on \gls{gpumd} \cite{Xu2025GPUMD} and \gls{nep} \cite{fan2021neuroevolution}. The same paradigm has been successfully validated in other systems, such as liquid water with NEP-MB-pol \cite{Xu2025NEPMBPOL}.
The framework combines the \gls{dft}-level accuracy characteristic of \gls{nep} with high speed and broad elemental coverage, enabling efficient simulations of complex materials and systems.
To ensure broad coverage of device-relevant configurations for NEP-FET, training data were drawn from the structurally diverse OMat24. Since OMat24 contains limited explicit interfacial structures, additional interfacial data were incorporated, generated via \gls{aimd} and \gls{al} simulations (see Methods and Supplementary Materials S1, S2 for details).
We use the enriched dataset to train \gls{nep} models tailored for \gls{fet} systems (see Methods and Supplementary Materials S3 for details). Following model training, device geometries are constructed using our FETMOD module, after which simulations are performed in \gls{gpumd} to evaluate the relevant material properties and device-level responses. To demonstrate the workflow, we consider a \gls{3d} silicon-based \gls{fet}, adapted from Refs.~\cite{Wang2015CONF, Sheng2024IEEETED}. This example serves as a representative test case for benchmarking the efficiency and predictive capability of the NEP-FET framework. The overall pipeline is illustrated in \autoref{fig:framework}a.

\begin{figure*}[ht]
\centering
\includegraphics[width=2\columnwidth]{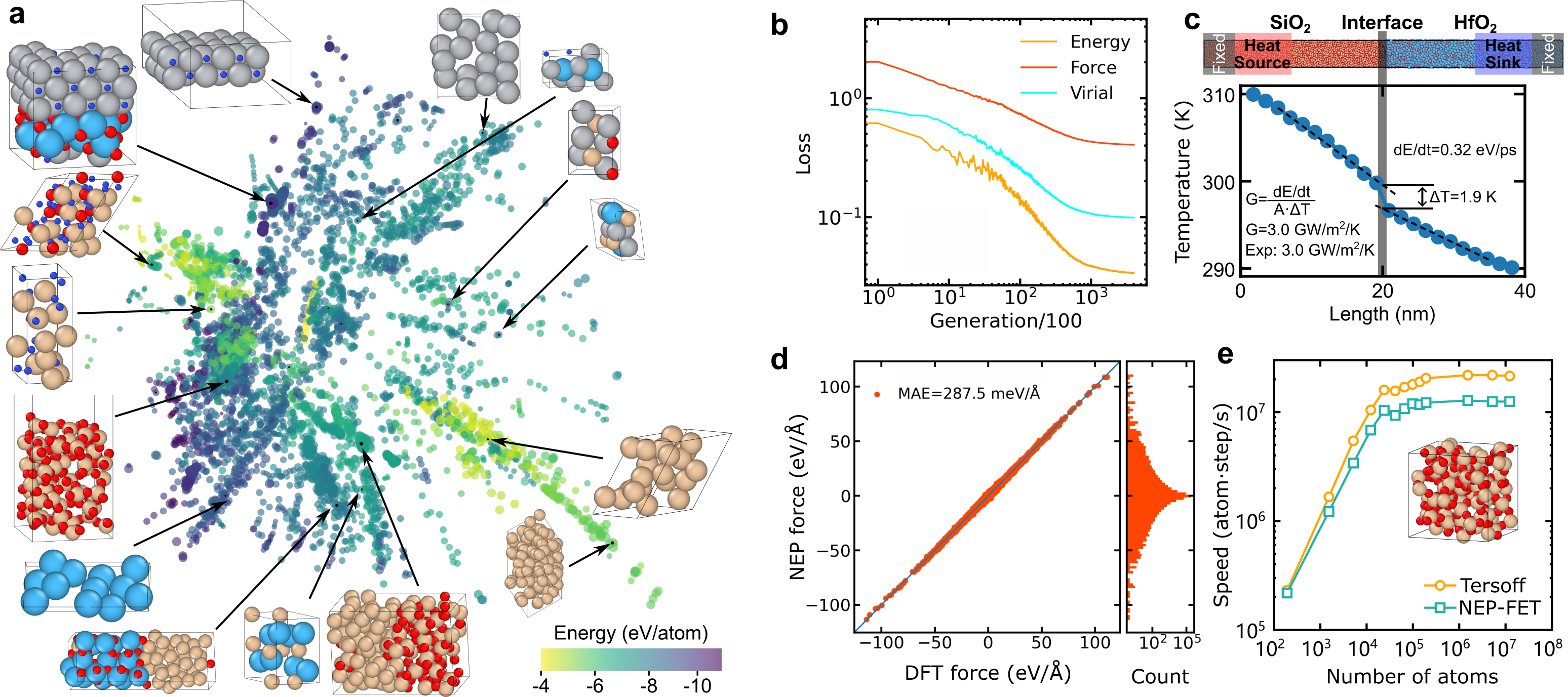}
\caption{\textbf{Dataset composition for NEP-FET and training accuracy.} 
(\textbf{a}) Distribution of the NEP-FET datasets in the reduced descriptor space spanned by the first two principal components. The color intensity indicates the average per-atom energy (in units of eV/atom) in the NEP-FET dataset, while the dot size represents the number of atoms contained in each individual training configuration. \textbf{b} Evolution of the training loss for the NEP-FET model. (\textbf{c}) shows the prediction of the interfacial heat transport properties of SiO$_2$ and HfO$_2$ materials by the NEP-FET model, demonstrating that the model can accurately capture the interfacial heat transport properties of the system \cite{chen2017aip}.  (\textbf{d}) Final force-prediction accuracy of the force prediction (\gls{mae} = 287.5 meV \AA$^{-1}$). (\textbf{e}) shows the benchmark results of computational efficiency of NEP-FET and traditional empirical potential model (Tersoff-SiO$_2$ \cite{Shinji2007CMS}) in SiO$_2$ system.}
\label{fig:nepmodel}
\end{figure*}

\noindent\textbf{FET device models:}
We compare two distinct device architectures: Fin\gls{fet} and the emerging \gls{gaa}\gls{fet}, both of which are shown in \autoref{fig:framework}b,c, with a particular focus on the cross-sectional view of the \gls{gaa}\gls{fet}. The device contains five representative interfaces: Si/SiO$_2$, Si/HfO$_2$, HfO$_2$/SiO$_2$, SiO$_2$/TiN, and HfO$_2$/TiN. The gate electrode is composed of titanium nitride (TiN), while the drain, source, and substrate are constructed from crystalline silicon (Si). \gls{sti} and other dielectric layers are SiO$_2$, with hafnium dioxide (HfO$_2$) used for the fin sidewall dielectric. For the Fin\gls{fet}, the source width $W1$ is 8{ }nm, the fin height $H4$ is 18{ }nm, the HfO$_2$ shell has a thickness of 1{ }nm, the gate width is 14{ }nm, and the gate-to-substrate distance is 2{ }nm. In the \gls{gaa} device, each channel (fin) has a height of 4{ }nm and a pitch of 5{ }nm. The simulation domain has dimensions of $35\times22\times35${ }nm$^3$ and contains approximately 2.0$\times10^{6}$ atoms. All models were built with our FETMOD module; additional sizes were generated by preserving these aspect ratios (see Methods and Supplementary Materials S4 for details).

\noindent\textbf{\gls{nep} model:} 
\gls{mlp}s trained on first-principles data have become a key component of modern materials-by-design workflows, as they can be adapted to specific classes of materials while retaining broad transferability. Among the available \gls{mlp} architectures, the \gls{nep} family achieves an attractive compromise between accuracy, speed, and accessible system size, and has been successfully applied to a wide variety of materials \cite{Xu2025NEPMBPOL,Ting2025NEP89}. In this work, we therefore employ \gls{nep} as our primary \gls{mlp} framework.

The \gls{nep} formalism \cite{fan2021neuroevolution} represents the total energy as a sum of atomic contributions, $E = \sum_i U_i$, where each site energy $U_i$ is obtained from a neural network that takes as input a set of local descriptors of the atomic environment. These descriptors are constructed from radial and angular basis functions built using Chebyshev and Legendre polynomials \cite{fan2021neuroevolution}, and are confined to a finite cutoff radius, so that the computational effort grows linearly with the number of atoms. The components of the descriptor vector serve as the input neurons of the network, and a single output neuron returns the scalar site energy $U_i$ for the central atom $i$. The network parameters are optimized using a regularized training procedure based on the separable natural evolution strategy \cite{schaul2011high}.
Forces and virial stresses are computed from the trained potential by taking analytical derivatives of the total energy with respect to atomic positions and strain. Chemical species are encoded through the expansion coefficients associated with a set of radial basis functions: for each ordered pair of species, the model introduces an independent set of $N_{\rm ec}$ expansion coefficients that are optimized during training, and each species is further assigned its own set of $N_{\rm nn}$ trainable neural network weights and biases. Detailed \gls{nep} parameters are provided in the Methods section and Supporting Information.

\noindent\textbf{Dataset construction and coverage:} Starting from OMat24, which contains periodic bulk structures for many semiconductors \cite{omat24}. Then, we applied farthest-point sampling to obtain a compact subset. Because OMat24 contains only periodic bulk structures and thus few explicit interfaces, we complemented this dataset with additional device-relevant environments: six classes of hetero-junctions and five types of surface structures. Representative members of these families were sampled by \gls{aimd} and active-learning \gls{md}, additional interface and surface configurations after \gls{dft} labeling. \autoref{fig:nepmodel}a illustrates the overall dataset coverage: we perform \gls{pca} on the local-environment descriptors to visualize the distribution of atomic environments across structures. Point colors indicate the mean atomic energy for each configuration, demonstrating both diversity and continuity of local motifs. Typical device interfaces represented in the dataset correspond to those in \autoref{fig:framework}b.

\begin{figure*}[ht]
\centering
\includegraphics[width=1.8\columnwidth]{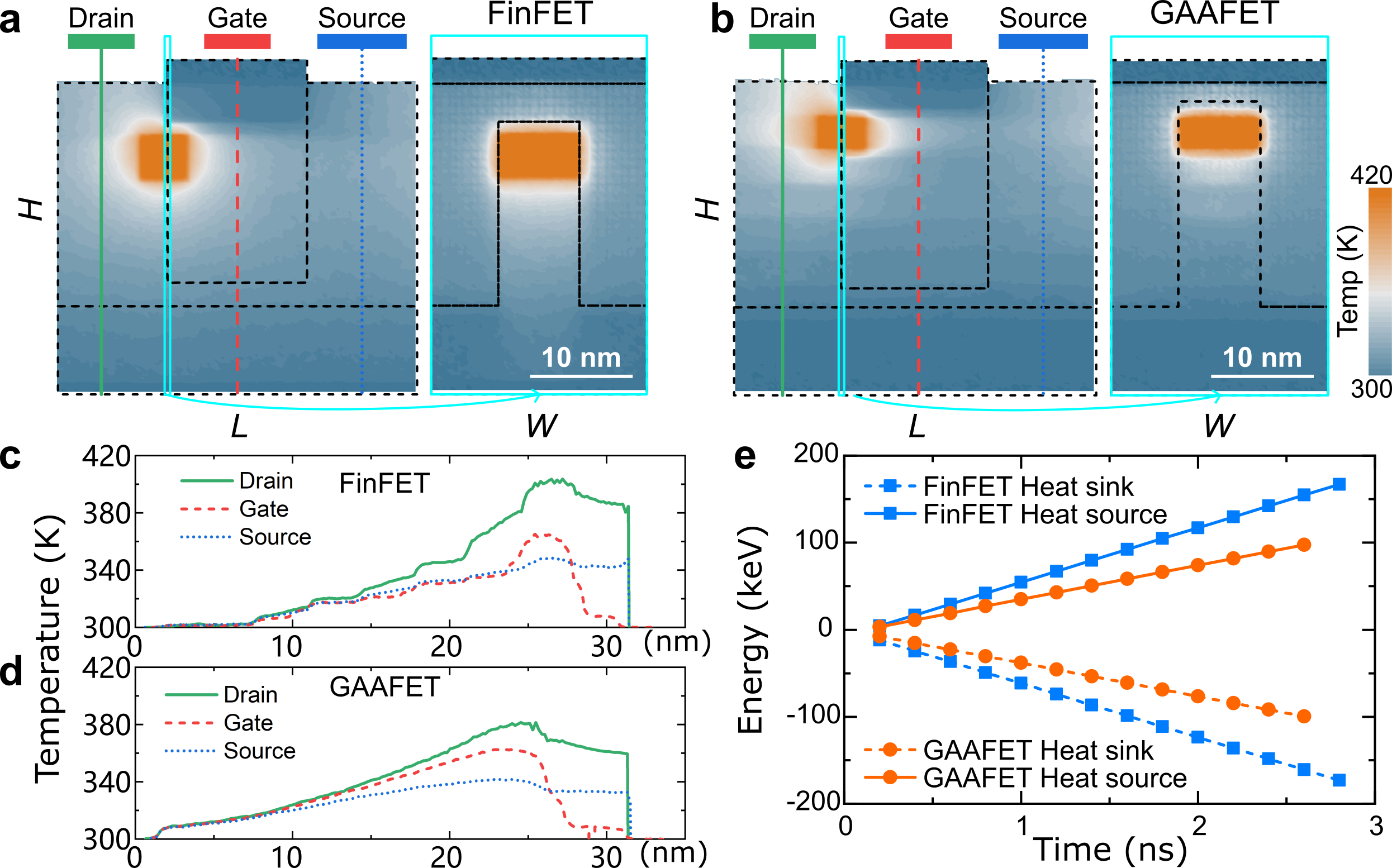}
\caption{
\textbf{Thermal-property model of NEP-FET devices.}
(\textbf{a}, \textbf{b}) Steady-state temperature maps for Fin\gls{fet} and \gls{gaa}\gls{fet}, respectively. Heat-source locations are defined in \autoref{fig:framework}b,d. The sections are taken normal to the two device interfaces; light-blue boxes delineate the regions from which the special cross-sections are extracted. Color encodes local temperature (see color bar). Green solid, red dashed, and blue dotted traces label the source, gate and drain regions, respectively. Position-dependent temperature profiles are plotted for (\textbf{c}) Fin\gls{fet} and (\textbf{d}) \gls{gaa}\gls{fet}. (\textbf{e}) Heat-flow distributions in the heat-source and heat-sink regions for both devices.}
\label{fig:tempprof}
\end{figure*}

\noindent\textbf{Model accuracy and validation:} The trained \gls{nep} model for the \gls{fet} systems (training protocol and hyperparameters in Methods and Supplementary Materials S1) achieves \gls{mae} relative to \gls{dft} references of 25.1 meV atom$^{-1}$ (energies), 287.5 meV\ \AA$^{-1}$ (forces), and 0.687 GPa (stresses) (\autoref{fig:nepmodel}b). \autoref{fig:nepmodel}d shows the parity plot of the forces from \gls{dft} and \gls{nep}, while the corresponding parity plots for the energy, virials, and stress are provided in the Supplementary Materials S6. These metrics are well within the accuracy range required for capturing anharmonic lattice dynamics and interfacial phonon scattering, key ingredients for predicting heat transport in disordered dielectric environments. To further validate the model’s transferability beyond the training set, we benchmarked the NEP-FET model on the \gls{itc} between HfO$_2$ and SiO$_2$, a technologically relevant oxide-oxide interface (see Methods for details). As shown in \autoref{fig:nepmodel}c, the predicted thermal conductance agrees closely with previous work \cite{chen2017aip}, demonstrating that the model not only reproduces static energy and force properties but also faithfully captures anharmonic interfacial heat transport. These results confirm that the NEP-FET model provides a reliable and computationally efficient surrogate for large-scale simulations of dielectric interfaces in advanced transistor architectures.

\noindent\textbf{Computational performance:} NEP-FET attains empirical-potential-like speed while retaining near-\gls{dft}-level fidelity. In \gls{md} of amorphous SiO$_2$ structures (\autoref{fig:nepmodel}e), the model sustains $1.25\times10^{7}$ atom-steps s$^{-1}$ on a single RTX{ }4090 card (\autoref{fig:nepmodel}e). For comparison, a Tersoff-SiO$_2$ \cite{Shinji2007CMS} potential achieves $2.14\times10^{7}$ atom-steps s$^{-1}$ on the same \gls{gpu}. NEP-FET thus attains empirical-potential-like speed despite its more complex many-body descriptors and larger effective cutoff. The resulting speed-accuracy balance enables systematic exploration of heat transport in realistic \gls{fet} geometries.

\begin{figure*}[ht]
\centering
\includegraphics[width=1.8\columnwidth]{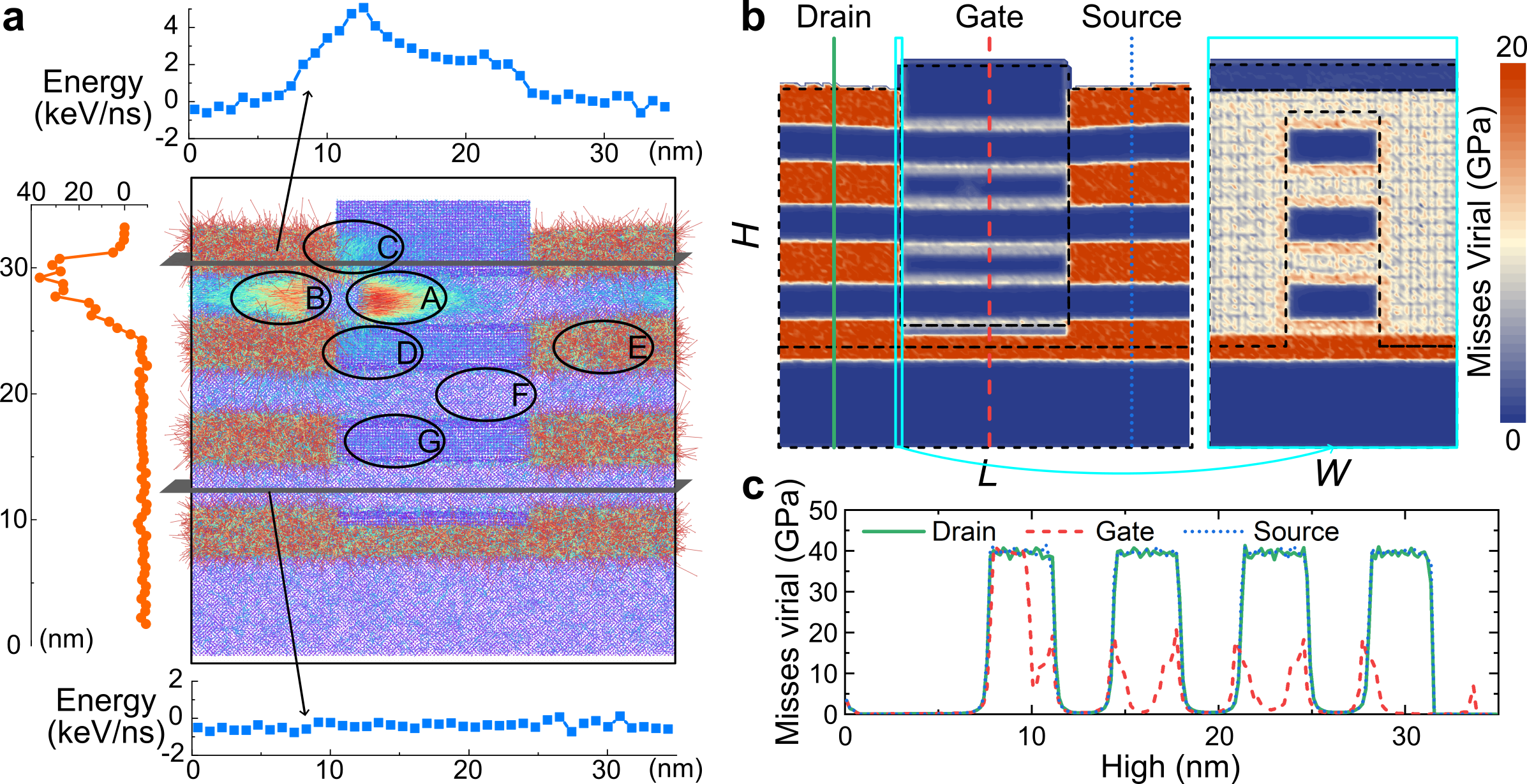}
\caption{
\textbf{Atomic-scale analysis of \gls{gaa}\gls{fet} device properties.}
(\textbf{a}) Per-atom heat-flux map of the device. Three dotted-line profiles show how the heat flux varies with position across the three regions: the left profile reports the global heat-flux-position relation for the entire device, whereas the upper and lower profiles give the corresponding relations for the grey-shaded subregions in the map.
(\textbf{b}) Per-atom von{ }Mises stress \cite{Wang1997Mises} distribution, with color indicating magnitude (see color bar).
(\textbf{c}) Von{ }Mises stress analysis of structures in different cross sections. Positional annotations within the corresponding regions use colors and line styles consistent with the profiles.}
\label{fig:atomic}
\end{figure*}

\noindent\textbf{Temperature maps:} Heat transport was evaluated using \gls{nemd}. Heat sources were positioned near the drain-side hotspot following finite-element pre-analysis predictions \cite{Hu2023JPCM,Hu2024FR,Sheng2024IEEETED}, with cold reservoirs defined at the top surface and at the substrate. Fixed boundary regions were introduced adjacent to the cold zones to ensure numerical stability and proper energy dissipation within the \gls{nemd} (see Methods for details). The resulting cross-sectional temperature distributions for the Fin\gls{fet} and \gls{gaa}\gls{fet} devices are shown in \autoref{fig:tempprof}a,b, together with temperature line profiles extracted along the indicated cutlines. In the \gls{gaa}\gls{fet}, heat remains tightly confined to the channel region and decays sharply away from the hotspot. Pronounced temperature discontinuities are observed across dielectric interfaces, signaling significant interfacial thermal resistance. In contrast, the Fin\gls{fet} shows a more laterally extended and uniform temperature field. The smoother gradients highlight more efficient heat spreading into the fin and stronger thermal coupling to the underlying substrate compared with the fully surrounded \gls{gaa} geometry.

\noindent\textbf{Effective thermal conductivity:} An effective thermal conductivity, $k_\mathrm{eff}$, was extracted once a steady-state temperature field was reached (procedure described in Methods). From \autoref{fig:tempprof}e, we obtain $k_\mathrm{eff} = 3.37~\mathrm{W{ }m^{-1}{ }K^{-1}}$ for the Fin\gls{fet} and $2.13~\mathrm{W{ }m^{-1}{ }K^{-1}}$ for the \gls{gaa} device, meaning that the Fin\gls{fet} exhibits a 58.2\% higher effective thermal conductivity. This enhancement arises from its larger proportion of high-conductivity crystalline Si along the primary heat-conduction paths and the improved structural continuity available for heat flow. Together, these factors enable more efficient thermal spreading and reduce the overall thermal resistance relative to the fully encapsulated \gls{gaa} architecture.

\noindent\textbf{Atomic heat flux analysis:} From per-atom heat fluxes (find the details in the Methods), we construct microscopic heat-flow maps across the device. In \autoref{fig:atomic} (labeled region A--F refer to the annotated regions), the \gls{gaa} device shows peak heat flux concentrated near the hotspot, with dominant lateral channels towards regions A and B, especially towards the gate-proximal side (region A), where the TiN interconnects assist heat removal (region C). Region D contributes most of the downward flux into the substrate. Region (E), composed of amorphous SiO$_2$, exhibits strong phonon scattering; heat leakage through this path is minimal compared with crystalline Si (region F) and TiN (regions D and G). The left inset of \autoref{fig:atomic}a quantifies the $z$-directed flux beneath the hotspot. The substrate-directed component is substantially smaller in the \gls{gaa} device ($\sim8.52$ keV ns$^{-1}$, $\sim22.3\%$ of total) than in the Fin\gls{fet} (versus $\sim39.1$ keV ns$^{-1}$, $\sim6 2.5\%$), consistent with nongray \gls{bte} trends \cite{Sheng2024IEEETED}. Consequently, in our \gls{gaa} model, the majority of heat dissipates laterally through the gate region, which motivates the use of high-thermal-conductivity gate stacks. In contrast, the Fin\gls{fet} exhibits a more substrate-dominated dissipation pathway, with pronounced heat flow from the fin base into the underlying Si and \gls{sti} regions, leading to a deeper and more vertically extended hotspot (see Supplementary Materials Fig. S7 for details). The upper inset of \autoref{fig:atomic}a shows the flux across the shaded gate section: a large fraction is carried by TiN, while SiO$_2$ contributes negligibly. The lower inset of \autoref{fig:atomic}a reports the flux across the device core, revealing a comparatively uniform in-plane distribution away from interfaces. Analogous spatial patterns are observed in the Fin\gls{fet}, but with a larger substrate-coupled fraction and a broader hotspot footprint along the fin sidewalls.

\noindent\textbf{Thermal stress:} \ref{fig:atomic}b presents the atomic von{ }Mises stress \cite{Wang1997Mises} in the \gls{gaa} device (see Methods for details), highlighting thermo-mechanical responses near the hotspot. Amorphous SiO$_2$ sustains the highest stresses, indicating significant built-in strain; HfO$_2$ near the gate also shows pronounced stress accumulation, whereas crystalline Si remains comparatively low, consistent with its role as a structural backbone. The blue boxes in \ref{fig:atomic}b mark cross-sectional stress maps through the highlighted regions, and the accompanying line plots compare stress across the drain, gate, and source. Stresses are consistently more concentrated in the gate region and peak near the hotspot, underscoring coupled thermal and mechanical reliability constraints.

\section{Conclusions \label{section:Conclusions}}

We have introduced NEP-FET, a computational framework that bridges the critical gap between quantum-mechanical accuracy and device-scale thermal simulations for advanced transistors. By leveraging a targeted \gls{mlp} trained on a compact yet representative dataset, this approach enables fully atomistic simulations of realistic \gls{fet} geometries, encompassing millions of atoms, with \gls{dft}-level fidelity on a single \gls{gpu}.

Our framework provides direct access to previously inaccessible atomistic fields, including spatially resolved temperature maps, per-atom heat flux, and thermal stress distributions. Applying NEP-FET to comparative studies of Fin\gls{fet} and \gls{gaa} architectures reveals fundamental insights into nanoscale heat dissipation. We find that the \gls{gaa} transistor exhibits more localized thermal hotspots and a significantly lower effective thermal conductivity compared to the Fin\gls{fet}, a consequence of its more confined architecture and greater interfacial density, which impedes efficient heat spreading towards the substrate. Atomic-scale analysis further elucidates the dominant heat-flow pathways and identifies regions of pronounced thermomechanical stress, particularly within amorphous gate dielectrics.

NEP-FET thus establishes a systematic and predictive pathway for investigating phonon-mediated heat transport and its coupling with mechanical response in advanced electronic devices. This capability is pivotal for addressing the escalating challenge of self-heating and for guiding the thermally aware design of future high-performance transistors.

\section{Methods \label{section:Methods}}

This workflow combines a curated, interface-rich training database, \gls{dft} labeling, a fine-tuned \gls{nep} model, and large-scale \gls{nemd} on device geometries generated with the FETMOD module. The framework yields spatially resolved temperature, heat flux and stress fields, from which effective thermal conductivity and thermo-mechanical reliability metrics are extracted.

\vspace{0.6em}

\noindent\textbf{Training datasets.}
The devices studied here involve five atomic species (O, Si, Hf, Ti and N). We first extracted 28{ }280 structures containing only these elements from the OMat24 dataset \cite{omat24}. A preliminary \gls{nep} model was fitted to this raw subset and used in a two-step “train-predict-clean” protocol: for each structure we compared NEP-predicted and reference \gls{dft} energies and removed outliers with large discrepancies. After two iterations, this yielded 22{ }721 bulk configurations with consistent \gls{dft} and \gls{nep} energies. To obtain a compact and diverse bulk subset, we then applied \gls{fps} in descriptor space, selecting 4{ }000 representative bulk structures for model selection and hyperparameter tuning (the results for \gls{pca} are shown in Supplementary Materials S5). Because OMat24 \cite{omat24} mainly contains periodic bulk phases and thus lacks explicit device-like non-bulk environments, we constructed a supplementary interface/surface dataset comprising six heterointerfaces (a-HfO$_2$/TiN, a-SiO$_2$/a-HfO$_2$, a-SiO$_2$/TiN, a-SiO$_2$/a-Si, a-Si/SiO$_2$, and a-Si/a-HfO$_2$) and five surface families (a-HfO$_2$, a-Si, a-SiO$_2$, TiN surfaces, and TiN nanowires), for a total of 11 classes of device-relevant local environments. Starting from the general-purpose NEP89 model \cite{Ting2025NEP89}, we fine-tuned a five-element \gls{nep} and used it to run NPT \gls{md} trajectories (0.5{ }fs time step, 1.5{ }ns length) for 11 representative interface/surface structures. For each trajectory, 50 frames were sampled at equal time intervals to reduce temporal correlations, and single-point \gls{dft} calculations (energies, forces and virials) were performed for all 550 configurations. After discarding a small number of pathological cases and adding a few additional structures from targeted melt-quench tests, this yielded 616 interface and surface configurations with \gls{dft} labels. Finally, the bulk set of 4{ }000 configurations and the 616 supplementary interface/surface configurations were merged, followed by a second \gls{fps} step to define a compact training core. This yielded a final set of 4{ }616 fully validated configurations used for NEP-FET training, consistent with the dataset statistics reported in the main text.

\vspace{0.6em}
\noindent\textbf{\gls{dft} calculations.}
All reference single-point \gls{dft} calculations were carried out with the {VASP} package (version 6.3.0) using the projector-augmented-wave (PAW) method \cite{Blochl1994PRB}. Exchange-correlation effects were treated within the generalized-gradient approximation using the \gls{pbe} functional \cite{Perdew1996PRL,Mohr2007PRB}. A plane-wave kinetic-energy cutoff of 500{ }eV (\texttt{ENCUT = 500}) and the “Accurate” precision setting (\texttt{PREC = Accurate}) were employed to suppress aliasing errors. Brillouin-zone integrations used the automatic $k$-point generation scheme (\texttt{KSPACING = 0.2, KGAMMA = .TRUE.}), corresponding to a reciprocal-space sampling density of $\sim$0.2{ }\AA$^{-1}$. Electronic self-consistency was converged to $10^{-6}${ }eV in total energy (\texttt{EDIFF = $1\times10^{-6}$}) and a maximum of 150 electronic iterations (\texttt{NELM = 150}) was allowed, which proved sufficient for robust convergence across the chemically and structurally diverse configurations in the training database. All calculations were performed in a non-spin-polarized setup (\texttt{ISPIN = 1}), with fixed ionic positions (\texttt{NSW = 1, IBRION = -1}). Partial occupancies were treated using the tetrahedron method with Blöchl corrections (\texttt{ISMEAR = -5}). Additional technical details of the reference data calculations are provided in Supplemental Notes S2 and S3.

\vspace{0.6em}
\noindent\textbf{\gls{nep} training.}
The \gls{nep} model for the five-component FET material system was trained using the fourth-generation \gls{nep} (NEP4) scheme \cite{song-nc-2024,fan2021neuroevolution,Xu2025GPUMD}. Following the standard \gls{nep} formalism, the total energy is expressed as a sum of site energies $U_i$ depending on atom-centered descriptors within a finite cutoff radius. In practice, we used a radial cutoff of 6.0{ }\AA{} and an angular cutoff of 5.0{ }\AA{} (\texttt{cutoff 6 5}), Chebyshev expansion orders $n_{\max}=4$ for both radial and angular channels with 8 basis functions per channel (\texttt{n\_max 4 4}, \texttt{basis\_size 8 8}), and angular-momentum limits $l_{\max}=(4,2,1)$ for the multi-body channels (\texttt{l\_max 4 2 1}). Short-range nuclear overlap was regularized by adding a Ziegler-Biersack-Littmark repulsive term smoothly truncated at 2{ }\AA{} (\texttt{zbl 2}). (see Supplementary Materials S3 for details)

The atomic energy mapping was implemented as a fully connected feedforward neural network with a single hidden layer of 80 neurons (\texttt{neuron 80}) and hyperbolic-tangent activation. Energies, forces and virial stresses were fitted simultaneously by minimizing a composite loss function defined as a weighted sum of the root-mean-square errors (RMSEs) of energies (per atom), forces and virials, plus $\mathcal{L}2$ regularization on the network parameters. We set equal weights for energies, forces and virials ($\lambda_\mathrm{e} = \lambda_\mathrm{f} = \lambda_\mathrm{v} = 1$; \texttt{lambda\_e 1}, \texttt{lambda\_f 1}, \texttt{lambda\_v 1}), retaining the default $\mathcal{L}2$ coefficient. Optimization used the separable natural evolution strategy built into NEP4 with a population size of 50 candidate parameter vectors per generation (\texttt{population 80}), and a maximum of $4\times10^5$ generations (\texttt{generation 200000}). The final NEP-FET model was chosen as the parameter set with the lowest validation loss and attains the energy, force and stress errors reported in \autoref{fig:nepmodel}.

\vspace{0.6em}
\noindent\textbf{Device construction and NEMD simulations.}
Atomistic Fin\gls{fet} and \gls{gaa}\gls{fet} geometries were generated using the FETMOD module. The gate metal was TiN, while the drain, source and substrate were crystalline Si. \gls{sti} and spacer layers were SiO$_2$, and the fin sidewall dielectric was HfO$_2$. For the Fin\gls{fet}, the source width ($D_\mathrm{W}$) was 8{ }nm, the fin height 18{ }nm, the HfO$_2$ shell thickness 1{ }nm, the gate width 14{ }nm and the gate-substrate spacing 2{ }nm. For the \gls{gaa} device, each channel (Fin) had a height of 4{ }nm and a pitch of 5{ }nm. The primary simulation box spanned $35\times22\times35${ }nm$^3$ and contained approximately $2.0\times10^{6}$ atoms; additional sizes were generated by scaling while preserving aspect ratios (see Supplementary S4 for details). Si substrates were oriented along (001) with channels aligned to $\langle 110\rangle$.

Amorphous oxide blocks were generated via melt-quench \gls{md} and then relaxed using the NEP-FET potential. These pre-relaxed slabs were joined to the crystalline Si and TiN regions with 0.3-0.5{ }nm interfacial relaxation windows; residual atomic overlaps were removed by soft-core pushes followed by energy minimization. For selected individual interfaces (e.g., Si/SiO$_2$, SiO$_2$/HfO$_2$, HfO$_2$/TiN), we also built simplified hetero-structure cells to benchmark \gls{itc} against reference calculations (Sec.{ }\ref{section:Results}).

All \gls{md} simulations were performed with \gls{gpumd} \cite{Xu2025GPUMD} using the NEP-FET model. Periodic boundary conditions were applied in the lateral directions. Along the transport direction, we included fixed boundary layers and Langevin thermostat regions to set up non-equilibrium heat flow. Each assembled device structure was first equilibrated in the isothermal-isobaric (NPT) ensemble at the target temperature and zero external pressure for 0.5{ }ns, followed by 0.25{ }ns of canonical (NVT) equilibration. The barostat was applied only along periodic directions so as not to distort free surfaces. The time step was 0.5{ }fs, which we verified to be sufficient for stable energy conservation and accurate heat-flux evaluation.

In the production NEMD runs (4.0{ }ns), two spatially separated local Langevin thermostats \cite{Li2019JCP} defined a hot and a cold region along the cross-plane direction, imposing a temperature difference of 60{ }K. This setup generated a steady non-equilibrium state with approximately constant heat flux across the device. For simplified interface cells, the corresponding temperature drop $\Delta T$ across the interface was used to evaluate the \gls{itc} $G$ via
\begin{equation}
\label{equation:NEMD-G}
G = \frac{\mathrm{d}E/\mathrm{d}t}{A \Delta T},
\end{equation}
where $S$ is the interfacial cross-sectional area and $\mathrm{d}E/\mathrm{d}t$ is the time-averaged energy exchange rate between thermostats and heat baths.

\vspace{0.6em}
\noindent\textbf{Temperature fields and effective thermal conductivity.}
Local temperature fields were obtained from atomic kinetic energies after removing bin-wise drift. The simulation cell was voxelized into cubic bins of edge length 0.5{ }nm. For each bin centered at position $\mathbf{r}$, the instantaneous kinetic temperature was computed from the atomic kinetic energies in that bin after subtracting the local center-of-mass velocity. Bins that overlapped with the Langevin thermostat regions were excluded from subsequent analyses.

Temperature maps were time-averaged over the steady-state window (typically 20 snapshots uniformly sampled over the last 4{ }ns). For visualization, the fields were lightly smoothed using a Gaussian kernel with width $\sigma=0.5${ }nm; all quantitative values reported in the main text were extracted from the raw data. One-dimensional temperature profiles were obtained along annotated cutlines through the device. Interfacial temperature jumps were estimated by linear fits to bulk-like regions on either side of the interface, extrapolated to the nominal interface plane.

The effective thermal conductivity along the transport direction was computed as
\begin{equation}
\kappa_\mathrm{eff} = \frac{Q L}{\Delta T},
\end{equation}
where $Q$ is the steady heat flux (net thermostat power divided by the cross-sectional area normal to the transport direction), $L$ is the separation between the centroids of the hot and cold slabs, and $\Delta T$ is the fitted temperature difference between these slabs. The thermal boundary conductance of a given interface was similarly obtained as $G = J / \Delta T_\mathrm{int}$, where $\Delta T_\mathrm{int}$ is the fitted temperature jump across that interface.

\vspace{0.6em}
\noindent\textbf{Microscopic heat flux.}
Microscopic heat current for atom $i$ is evaluated using the following definition \cite{Xu2025GPUMD},
\begin{equation}
\mathbf{J}_i = \sum_{j \neq i}\mathbf{r}_{ij}
\left(\frac{\partial U_j}{\partial \mathbf{r}_{ji}}\cdot \mathbf{v}_i\right),
\end{equation}
For a group of atoms within a volume $V$, the heat flux is
\begin{equation}
\mathbf{Q} = \frac{1}{V} \sum_{i}\mathbf{J}_i,
\end{equation}
where $U_j$ is the site energy of atom $j$, $\mathbf{v}_i$ is the velocity of atom $i$, and $\mathbf{r}_{ij} = \mathbf{r}_j - \mathbf{r}_i$. The \gls{gpumd} implementation partitions many-body contributions in a manner consistent with the \gls{nep} energy decomposition. Local heat-flux vectors were coarse-grained on the same voxel grid as the temperature fields by summing $\mathbf{Q}_i$ over atoms in each bin and dividing by the bin volume, and then time-averaged over the steady-state window. The one-dimensional heat-flux profiles shown in the main text correspond to area averages over transverse directions, retaining only the component along the transport direction.

\vspace{0.6em}
\noindent\textbf{Stress fields and von{ }Mises stress.}
Local stresses were obtained from the microscopic virial tensor. Assuming Newton’s third law holds, the per-atom virial is \cite{Xu2025GPUMD}
\begin{equation}
\boldsymbol{W}_i = \sum_{j\neq i} \mathbf{r}_{ij} \otimes \frac{\partial U_j}{\partial \mathbf{r}_{ji}}.
\end{equation}
The Cauchy stress tensor within a voxel of volume $V_\text{bin}$ was then defined as
\begin{equation}
\boldsymbol{\sigma} = -\frac{1}{V_\text{bin}}\sum_{i\in \text{bin}} \boldsymbol{W}_i.
\end{equation}
Voxel-level stress tensors were time-averaged over the steady-state portion of the trajectory to obtain smooth continuum-like stress fields.

From the stress components, we computed the von{ }Mises stress as \cite{Wang1997Mises}
\begin{align}
\sigma_\mathrm{vM}
&= \frac{1}{\sqrt{2}}
\Bigg[
\frac{
(\sigma_{xx}-\sigma_{yy})^2
+ (\sigma_{yy}-\sigma_{zz})^2
+ (\sigma_{zz}-\sigma_{xx})^2}{2}
\notag\\[4pt]
&\qquad
+\,6\left( \sigma_{xy}^2 + \sigma_{yz}^2 + \sigma_{zx}^2 \right)
\Bigg]^{1/2}.
\end{align}
To isolate thermo-elastic contributions, we report stress differences relative to an isothermal 300{ }K reference simulation performed under identical boundary conditions, excluding bins that overlap with the thermostats.

\vspace{0.5cm}
\noindent{\textbf{Data availability:}}

The training and test datasets and the trained machine-learned potential models will be made available upon publication.

\vspace{0.5cm}
\noindent{\textbf{Code availability:}}

The source code and documentation for \gls{gpumd} are available
at \url{https://github.com/brucefan1983/GPUMD} and \url{https://gpumd.org}, respectively. The source code for FETMOD is available at \url{https://github.com/Kick-H/FETMOD}.

\vspace{0.5cm}
\noindent{\textbf{Declaration of competing interest:}}

The authors declare that they have no competing interests.

\section*{Contributions}
KX proposed the initial idea, constructed the dataset, trained the \gls{nep} model, and performed all the \gls{md} calculations.
KX and TL performed the benchmark tests.
YX and GW carried out the density functional theory calculations.
DD, HG, and XG provided partial computational support.
XK, LT, XW, GZ, and JX drafted the manuscript.
All authors discussed the results and contributed to the writing of the manuscript.

\begin{acknowledgments}

This work was supported by the Department of Science and Technology of Liaoning Province (No. JYTMS20231613). 


\end{acknowledgments}

\bibliographystyle{myaps}
\bibliography{ref.bib}
\end{document}